# GradMix for nuclei segmentation and classification in imbalanced pathology image datasets

Tan Nhu Nhat Doan[1], Kyungeun Kim[2], Boram Song[2], and Jin Tae Kwak[3]

[1]Department of Computer Science and Engineering, Sejong University, Seoul 05006, Korea
[2]Department of Pathology, Kangbuk Samsung Hospital, Sungkyunkwan University School of Medicine, Seoul 05505, Korea
[3]School of Electrical Engineering, Korea University, Seoul 02841, Korea
`jkwak@korea.ac.kr`

**Abstract.** An automated segmentation and classification of nuclei is an essential task in digital pathology. The current deep learning-based approaches require a vast amount of annotated datasets by pathologists. However, the existing datasets are imbalanced among different types of nuclei in general, leading to a substantial performance degradation. In this paper, we propose a simple but effective data augmentation technique, termed GradMix, that is specifically designed for nuclei segmentation and classification. GradMix takes a pair of a major-class nucleus and a rare-class nucleus, creates a customized mixing mask, and combines them using the mask to generate a new rare-class nucleus. As it combines two nuclei, GradMix considers both nuclei and the neighboring environment by using the customized mixing mask. This allows us to generate realistic rare-class nuclei with varying environments. We employed two datasets to evaluate the effectiveness of GradMix. The experimental results suggest that GradMix is able to improve the performance of nuclei segmentation and classification in imbalanced pathology image datasets.

## 1 Introduction

The assessment of nuclei is one of the primary tasks in digital pathology since nuclear features, including shape, size, and density, have known to be related to disease diagnosis and prognosis [1]. In order to analyze nuclear features in pathology images, an accurate and reliable segmentation and classification of nuclei is a pre-requisite. However, nuclei segmentation and classification remain a challenging task since there exists an enormous number of nuclei in a relatively small pathology image and there is a substantial intra- and inter-variability in the morphology, texture, and intensity among nuclei of differing cell types as well as within the same cell type. Hence, a robust nuclei segmentation and classification method is needed to expedite digital pathology analysis and to improve diagnostic decisions on pathology images.

Recently, several research efforts have made to develop deep learning-based nuclei segmentation and classification methods. Most of them focused on nuclei segmentation where one of the most challenging tasks is to separate touching or overlapping nuclei [2]. Some designed multi-resolution convolutional neural networks (CNNs) to preserve contextual information at multiple resolutions [3] [4]. Some others proposed to exploit morphology of nuclei. For example, [5] [4] utilized nuclear boundaries in identifying individual nuclei. [6] formulated nuclei segmentation as a regression of the distance map of nuclei. [7] exploited both the nuclear distance map and nuclear boundary for nuclei segmentation. Moreover, [8] utilized dense steerable filters to achieve rotation-symmetry within the network. Nuclei classification has been mainly studied as a downstream analysis of nuclei segmentation. For instance, [9] detected nuclear centroids and classified nuclei using CNN. [10] proposed RCCNet that classifies nuclei image patches into pertinent classes. [2] developed HoVer-Net that simultaneously performs nuclei segmentation and classification by utilizing horizontal and vertical distance maps of nuclei. Despite such recent advances, nuclei segmentation and classification still need to be improved. There exists a high variability in both segmentation and classification performance among different types of nuclei [2] [11]. This may be ascribable to imbalance in the datasets among different nuclei types. The lesser the annotated nuclei are available, the poorer the performance is in general. For nuclei segmentation, [12, 13] proposed to use a generative adversarial network (GAN) to synthesize pathology images with known nuclei annotations; however, GAN is not only computationally expensive but also requires a sufficient number of supervised datasets. Mixup [14], Cutout[15], and CutMix [16] are regularization techniques to generate new images from the existing images. These are computationally efficient and have been successfully applied to image classification and object detection. But, no prior work exists for image segmentation in digital pathology.

Herein, we propose a gradation mixup (GradMix) data augmentation technique for an improved nuclei segmentation and classification in imbalanced pathology image datasets. In the imbalanced datasets, there exist one or more major-classes of nuclei that are prevalent in the datasets and one or more rare-classes of nuclei that are under-represented in the datasets. GradMix is a data augmentation technique that is tailored to nuclei segmentation and classification tasks. The technique aims at increasing the number of rare-class nuclei by generating realistic nuclei under various environments. GradMix generates a new rare-class nucleus by combining a major-class nucleus with a rare-class nucleus via a customized mixing mask $\mathcal{M}$. The rare-class nucleus is utilized as it is and placed at the center of the major-class nucleus. Then, the neighboring pixels of the rare-class nucleus and the corresponding pixels that are either major-class nucleus or its neighbors are combined with the corresponding weights in $\mathcal{M}$. The weights for the neighboring pixels of the rare-class nucleus are inversely related to the distance to the boundary of the rare-class nucleus. In this manner, we are able to generate a set of new, realistic rare-class nuclei with varying environments. This, in turn, aids in improving the performance of nuclei segmentation and classification.

## 2  Methodology

The overview of the proposed GradMix is illustrated in **Fig. 1**. **Algorithm 1** provides the detailed algorithm for GradMix.

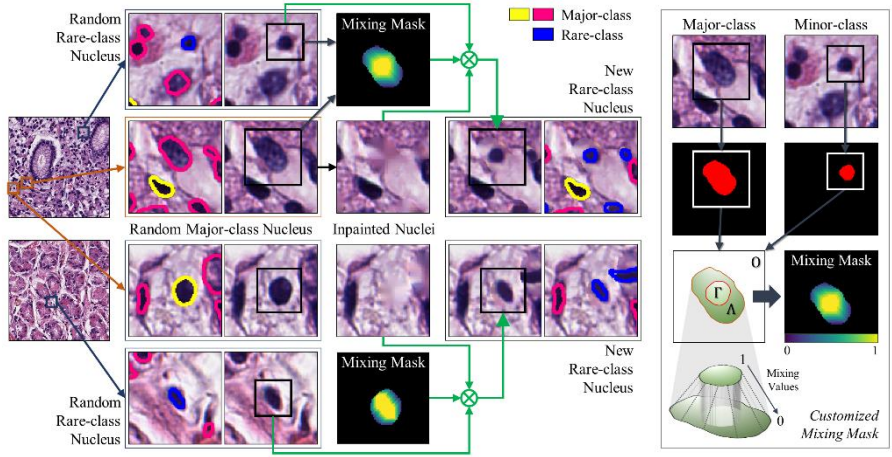

**Fig. 1.** Overview of GradMix. Major-class and rare-class nuclei are randomly selected and mixed to generate new rare-class nuclei.

### 2.1  GradMix

Let $\mathcal{J} = \{(I_i, G_i) | i = 1, ..., N\}$ be a set of a H&E image and ground truth map where $I_i$ is the $i$th H&E image, $G_i$ is the $i$th ground truth map, and $N$ is the number of image-ground truth map pairs. In an image $I_i$, there exist a number of major-class nuclei $\{x_{i,j}^m | j = 1, ..., N_i^m\}$ and a number of rare-class nuclei $\{x_{i,j}^r | j = 1, ..., N_i^r\}$ where $x_{i,j}^m \in \{1, ..., C^m\}$ and $x_{i,j}^r \in \{1, ..., C^r\}$ are the $j$th instances of major-class and rare-class nuclei, $C^m$ and $C^r$ are the cardinality of the major-class and rare-class nuclear types, and $N^m$ and $N^r$ are the number of major-class and rare-class nuclei, respectively. The objective of GradMix is to generate a new instance of rare-class nuclei $\hat{x}^r$ by combining one instance of major-class nuclei $x^m$ and one instance of rare-class nuclei $x^r$. To combine the two instances, we define a mixing function as follows:

$$\hat{x}^r = \mathcal{M} \odot \rho(x^r) + (1 - \mathcal{M}) \odot x^m \tag{1}$$

where $\mathcal{M} \in [0,1]$ is a customized mixing mask, $\odot$ is element-wise multiplication, and $\rho$ is an image inpainting function [17]. The image inpainting function $\rho(x)$ replaces the intensity of the pixels inside $x$ with the intensity that is similar to the neighboring pixels via interpolation, removing the color and texture of $x$.

To create the customized mixing mask $\mathcal{M}$ for the pair $(x^r, x^m)$, we first conduct morphological dilatation for $x^m$ with a 3x3 square kernel so as to include the neighboring pixels of $x^m$. The dilated instance is designated as $\tilde{x}^m$. $x^r$ is translated to match the centroid of $x^r$ to the centroid of $x^m$, forming $\tilde{x}^r$. Then, we define three mutually exclusive sets of pixels as $O = \{u | u \notin \tilde{x}^m\}$, $\Gamma = \{u | u \in \tilde{x}^r\}$, and $\Lambda = \{u | u \in \tilde{x}^m \text{ and } u \notin \tilde{x}^r\}$ where O, $\Gamma$, and $\Lambda$ denote the sets of pixels that are corresponding to the outside of $\tilde{x}^m$, inside of $\tilde{x}^r$, and inside of $\tilde{x}^m$ excluding $\tilde{x}^r$, respectively. For the sets O and $\Gamma$, a mixing value is assigned per pixel as follows: $\forall u \in O, \mathcal{M}(u) = 1$ and $\forall u \in \Gamma, \mathcal{M}(u) = 0$. For the set $\Lambda$, the mixing value is computed as a normalized minimum distance to $\tilde{x}^r$:

$$\mathcal{M}(u) = \frac{\min_{v \in \tilde{x}^r} D(u,v)}{\sum_{u' \in \Lambda} \min_{v \in \tilde{x}^r} D(u',v)}, \forall u \in \Lambda \quad (2)$$

where $D$ denotes Euclidean distance.

For each image $I_i$, we randomly select 80% of the major-class nuclei $\{x_{i,j}^m | j = 1, \dots, N_i^m\}$, combines each instance of the selected $x^m$ with a randomly selected instance of $x^r$ whose size is smaller than half of the size of $x^m$, and generates a set of new instances of rare-class nuclei $\tilde{x}^r$. As we select an instance of the rare-class nuclei, we consider all $x^r$ not only in the same H&E image but also in other H&E images to enhance the diversity of the rare-class nuclei. Specifically, those rare-class nuclei are randomly selected from $\{x_{i,j}^r | j = 1, \dots, N_i^r\}$ with a 60% chance and from $\{x_{k,j}^r | j = 1, \dots, N_k^r\}_{k=1, k \neq i}^N$ with a 40% chance. To adjust the color difference among differing images, we compute the difference in the average color intensities of all the nuclei in $I_i$ and the selected $x^r$. Then, the intensity difference is added to all the pixels in $x^r$.

## 2.2 Network Architecture & Optimization

Following [18], we build CNN for simultaneous nuclei segmentation and classification. The network contains one encoder and three decoders. The encoder has 2 convolution (Conv) blocks and 6 mobile inverted bottleneck convolution (MBConv) blocks with squeeze-and-excitation [19]. The three decoders, nuclear foreground (NF), nuclear ordinal regression (NO), and nuclear type (NT) decoders, share the identical structure that has three repetitions of an up-sampling block, a dilated convolution block with a factor of 1 and 2, and a 3×3 Conv block, followed by a series of a 5×5 Conv block, an up-sampling block, a 5×5 Conv block, and a 1×1 Conv block. NF and NT predict a nuclear foreground map and nuclear type map, respectively. NO predicts a nuclear ordinal regression map where each nucleus is subdivided into K layers with distinct ordinal labels. The ordinal labels are determined by the decreasing distance from the boundary to the centroid of the nucleus.

During training, the output of NO is utilized to obtain the two outermost layers (near boundary) that are corresponding to the most challenging regions since they determine the exact shape of nuclei. The two outermost layers are used to assign weights for NF and NT (2 for the outermost layer and 1 for the second outermost layer) so as to give more attention to the challenging regions at each iteration.

55

To optimize the network, we adopt cross-entropy loss for NO, cross-entropy loss and DICE loss for NF, and focal loss for NT. For NO, cross-entropy loss is computed for each ordinal layer. The weights of the two outermost layers are utilized for the computation of cross-entropy loss and focal loss for NF and NT, respectively.

---

**Algorithm 1** GradMix

**Inputs**: A set of images $\{I_i|i = 1, ..., N\}$ with a set of major-class nuclei $\{x_{i,j}^m|j = 1, ..., N_i^m\}$, a set of rare-class nuclei $\{x_{i,j}^r|j = 1, ..., N_i^r\}$, $\rho$: an inpainting function

**Outputs**: A set of new instances of rare-class nuclei $\{\hat{x}_{i,j}^r|j = 1, ..., N_i^{|\Phi|}\}$

1: **for** each image $I_i$ **do**
2:     Select 80% of $\{x_{i,j}^m|j = 1, ..., N_i^m\}$ at random, forming $\Phi$
3:     Compute $\mu_i \leftarrow Average(I_i(\Phi))$
4:     **for** each instance $x^m \in \Phi$ **do**
5:        Initialize $\mathcal{M} \leftarrow 0$
6:        Randomly select an instance $x^r$, subject to $size(x^r) < size(x^m)/2$, from $\{x_{i,j}^r|j = 1, ..., N_i^r\}$ and $\{x_{k,j}^r|j = 1, ..., N_k^r\}_{k=1, k \neq i}^N$ with 60% and 40% chances, respectively
7:        Compute $\mu^m \leftarrow Average(I_i(x^r))$
8:        Adjust color intensity: $x^r \leftarrow x^r + \mu^m - \mu_i$
9:        $\tilde{x}^m \leftarrow MorphologicalDilation(x^m)$ with a 3x3 square kernel
10:       $\tilde{x}^r \leftarrow Translation(x^r, x^m)$: Match the centroid of $x^r$ to that of $x^m$
11:       Generate O, $\Gamma$, and $\Lambda$: $O \leftarrow \{u|u \notin \tilde{x}^m\}$, $\Gamma \leftarrow \{u|u \in \tilde{x}^r\}$, $\Lambda \leftarrow \{u|u \in \tilde{x}^m \text{ and } u \notin \tilde{x}^r\}$
12:       Assign mixing values: $\forall u \in O, \mathcal{M}(u) \leftarrow 1$, $\forall u \in \Gamma, \mathcal{M}(u) \leftarrow 0$,
$$\forall u \in \Lambda, \mathcal{M}(u) \leftarrow \frac{\min_{v \in \tilde{x}^r} D(u,v)}{\sum_{u' \in \Lambda} \min_{v \in \tilde{x}^r} D(u',v)}$$
13:       Synthesize a new mixed rare-class nucleus: $\hat{x}^r \leftarrow \mathcal{M} \odot \rho(x^r) + (1 - \mathcal{M}) \odot x^m$

---

## 3 Experiments and Results

**Table 1.** Details of datasets.

| Dataset | | Lymphocytes/ Inflammatory | Epithelial | Miscellaneous | Spindle | Total |
|---|---|---|---|---|---|---|
| GLySAC | Train | 7409 | 7154 | 3386 | - | 17949 |
| | Train+GradMix | 9674 | 9383 | 16841 | - | 35898 |
| | Test | 4672 | 5133 | 3121 | - | 12926 |
| CoNSeP | Train | 3941 | 5537 | 371 | 5700 | 15549 |
| | Train+GradMix | 6911 | 8218 | 7234 | 8741 | 31098 |
| | Test | 1638 | 3214 | 561 | 3357 | 8770 |

### 3.1 Datasets

Two nuclei segmentation and classification datasets are employed in this study (**Table 1**). The GLySAC dataset [18] is composed of 59 H&E images of size 1000x1000 pixels that were obtained at 40x magnification from gastric adenocarcinoma WSIs (Aperio digital scanner). The dataset has 30875 nuclei that are grouped into three categories, including 12081 lymphocytes, 12287 epithelial nuclei, and 6507 miscellaneous. The miscellaneous category denotes any nuclei that do not belong to lymphocytes and epithelial nuclei categories such as stromal nuclei, endothelial nuclei, and etc. The 59 images are split into a training set (34 images) and a test set (25 images). The CoNSeP dataset [2] contains 24319 annotated nuclei from 41 H&E images of size 1000x1000 pixels. The nuclei are grouped into four types, including 5537 epithelial nuclei, 3941 inflammatory, 5700 spindle-shaped nuclei, and 371 miscellaneous. The 41 images are divided into a training stet of 27 images and a test set of 14 images.

### 3.2 Implementation Details

All the networks are implemented with the open-source software library Tensorflow version 1.12 on a workstation with two NVIDIA GeForce 2080 Ti GPUs. We train the networks for 50 epochs. During training, Adam optimizer is used with an initial learning rate of $1e^{-4}$, which is subsequently reduced to $1e^{-5}$ after 25 epochs. Several online data augmentation techniques are applied, including a random flip and rotation, a Gaussian blur, a Median blur, a Gaussian noise, and a random color change.

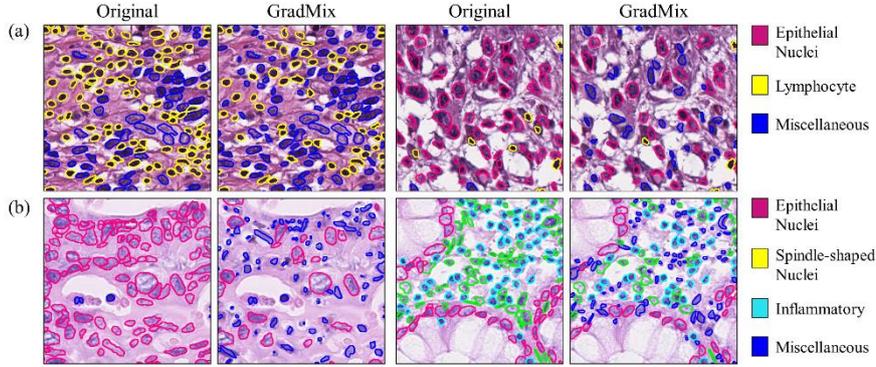

**Fig. 2.** Exemplary GradMix-ed images for (a) GLySAC and (b) CoNSeP datasets.

### 3.3 Results and Discussions

The simultaneous nuclei segmentation and classification network was separately evaluated on the two datasets (GLySAC and CoNSeP). To assess the effectiveness of GradMix, the network was trained on the combination of the original training set and GradMix-ed training set, and then tested on the test set of the two datasets. **Fig. 2** shows the exemplary input images and GradMix-ed images. Most of the major-class nuclei were replaced by the rare-class nuclei that were randomly chosen from the same and different images. The new mixed rare-class nuclei preserve the original texture and morphometry but possess a different neighboring environment. To quantitatively evaluate nuclei segmentation, dice coefficient (DICE), aggregated Jaccard Index (AJI), and panoptic quality (PQ) are utilized. PQ is composed of detection quality (DQ) and segmentation quality (SQ). As for nuclei classification, we quantified detection quality ($F_d$) and F1-score for each type of nuclei as described in [2]. In comparison to GradMix, we adopted CutMix [16] and repeated the same experiments by replacing the GradMix-ed images with the CutMix-ed images. Similar to GradMix, CutMix mixes two images to generate a new image. But, CutMix randomly samples a rectangular region in one image and replaces it by the corresponding region in the other image. To use CutMix, we randomly select $x^r$ and $x^m$ as described in **Algorithm 1** and replace $x^m$ by the rectangular region encompassing $x^r$.

**Table 2.** Results of nuclei segmentation.

|  | DICE | AJI | DQ | SQ | PQ |
|---|---|---|---|---|---|
| GLySAC | 0.838 | 0.672 | 0.811 | 0.796 | 0.648 |
| +CutMix | 0.833 | 0.660 | 0.803 | 0.791 | 0.637 |
| +GradMix (proposed) | **0.849** | **0.680** | **0.814** | **0.802** | **0.655** |
| CoNSeP | 0.844 | 0.586 | **0.698** | 0.772 | 0.540 |
| +CutMix | 0.842 | 0.586 | 0.693 | 0.772 | 0.537 |
| +GradMix (proposed) | **0.846** | **0.589** | **0.698** | **0.778** | **0.545** |

**Table 3.** Results of nuclei classification.

|  | $F_d$ | $F^E$ | $F^L$ | $F^M$ | $F^I$ | $F^S$ |
|---|---|---|---|---|---|---|
| GLySAC | 0.864 | **0.557** | 0.535 | 0.360 | - | - |
| +CutMix | 0.855 | 0.565 | 0.522 | 0.315 | - | - |
| +GradMix (proposed) | **0.872** | 0.556 | **0.543** | **0.395** | - | - |
| CoNSeP | 0.778 | 0.662 | - | 0.465 | **0.659** | 0.597 |
| +CutMix | 0.765 | 0.655 | - | 0.400 | 0.641 | 0.582 |
| +GradMix (proposed) | **0.780** | **0.678** | - | **0.476** | 0.651 | **0.598** |

$F^E$, $F^L$, $F^I$, $F^S$, and $F^M$ denote F1-score for epithelial nuclei, lymphocytes, inflammatory nuclei, spindle-shaped nuclei, and miscellaneous nuclei, respectively.

**Table 2** demonstrates the results of nuclei segmentation. For both GLySAC and CoNSeP datasets, the network equipped with GradMix was able to achieve the best segmentation performance, outperforming the ones using the original dataset only and the dataset with CutMix. The results of nuclei classification are available in **Table 3**. Similar to nuclei segmentation, we obtained the best classification performance by utilizing GradMix in general. The performance gain in nuclei classification is, in fact, (relatively) higher than that in nuclei segmentation. It is remarkable that we acquired a larger performance gain for rare-class nuclei, which are miscellaneous nuclei ($F^M$) for both datasets. With the help of GradMix, the F1-score for $F^M$ increased by 0.035 and 0.011 for GLySAC and CoNSeP datasets, respectively. These results on the two datasets suggest that the proposed GradMix could aid in improving both nuclei segmentation and classification performance, in particular for rare-class nuclei. Moreover, the performance gain in nuclei classification will benefit the downstream analysis of differing cell/nuclei types in pathology images, leading to an improved decision making on them.

The network with CutMix was unable to improve neither nuclei segmentation nor nuclei classification, even inferior to the one with the original dataset only. Since CutMix simply replaces one nucleus by the other using a rectangular window, there is a substantial discontinuity around the boundary of a new nucleus. In particular, for overlapping nuclei, it cuts out parts of neighboring nuclei. Such discontinuity and artifacts obviously have an adverse effect on nuclei segmentation and classification. The results with CutMix also indicate that the performance gain by GradMix was not solely due to the increased number of the major- or rare-class nuclei.

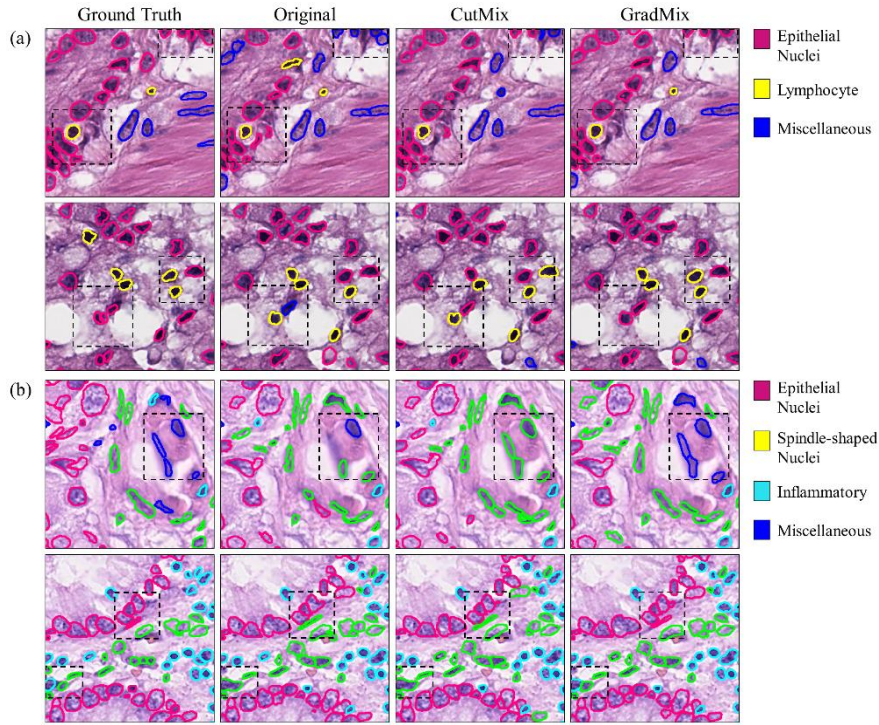

**Fig. 3.** Visual results for (a) GLySAC and (b) CoNSeP datasets.

**Fig. 3** demonstrates the visual results of nuclei segmentation and classification for the two datasets. Regardless of the datasets, the predictions by the network with GradMix was superior to others in handling overlapping nuclei as well as identifying the pertinent type of each instance of nuclei.

This study has several limitations. The experiments include a single model and two datasets. GradMix is only compared to CutMix. We will conduct an extended study to further validate GradMix by including more nuclei segmentation and classification models, external datasets, and data augmentation/generation techniques.

## 4    Conclusions

We present GradMix for an improved nuclei segmentation and classification in imbalanced pathology image datasets. GradMix is designed to mix the existing major-class nuclei and rare-class nuclei to generate new, mixed rare-class nuclei so as to increase the number of rare-class nuclei with varying conditions in an efficient and effective manner. The experimental results suggest that GradMix holds potential for resolving data imbalance issues in pathology image analysis.

**Acknowledgements**. This work was supported by the National Research Foundation of Korea (NRF) grant funded by the Korea government (MSIT) (No. NRF-2021R1A2C2014557 and No. NRF-2021R1A4A1031864).

## References

1. Li, X., Li, C., Rahaman, M.M., Sun, H., Li, X., Wu, J., Yao, Y., Grzegorzek, M.: A comprehensive review of computer-aided whole-slide image analysis: from datasets to feature extraction, segmentation, classification and detection approaches. Artificial Intelligence Review (2022)
2. Graham, S., Vu, Q.D., Raza, S.E.A., Azam, A., Tsang, Y.W., Kwak, J.T., Rajpoot, N.: Hover-net: Simultaneous segmentation and classification of nuclei in multi-tissue histology images. Medical Image Analysis 58, 101563 (2019)
3. Raza, S.E.A., Cheung, L., Shaban, M., Graham, S., Epstein, D., Pelengaris, S., Khan, M., Rajpoot, N.M.: Micro-Net: A unified model for segmentation of various objects in microscopy images. Medical image analysis 52, 160-173 (2019)
4. Vu, Q.D., Graham, S., Kurc, T., To, M.N.N., Shaban, M., Qaiser, T., Koohbanani, N.A., Khurram, S.A., Kalpathy-Cramer, J., Zhao, T.: Methods for segmentation and classification of digital microscopy tissue images. Frontiers in bioengineering and biotechnology 53 (2019)
5. Zhou, Y., Onder, O.F., Dou, Q., Tsougenis, E., Chen, H., Heng, P.-A.: Cia-net: Robust nuclei instance segmentation with contour-aware information aggregation. In: International Conference on Information Processing in Medical Imaging, pp. 682-693. Springer, (Year)
6. Naylor, P., Laé, M., Reyal, F., Walter, T.: Segmentation of nuclei in histopathology images by deep regression of the distance map. IEEE transactions on medical imaging 38, 448-459 (2018)
7. Liu, X., Guo, Z., Cao, J., Tang, J.: MDC-net: a new convolutional neural network for nucleus segmentation in histopathology images with distance maps and contour information. Computers in Biology and Medicine 135, 104543 (2021)
8. Graham, S., Epstein, D., Rajpoot, N.: Dense steerable filter cnns for exploiting rotational symmetry in histology images. IEEE Transactions on Medical Imaging 39, 4124-4136 (2020)

10
9. Sirinukunwattana, K., Raza, S.E.A., Tsang, Y.-W., Snead, D.R., Cree, I.A., Rajpoot, N.M.: Locality sensitive deep learning for detection and classification of nuclei in routine colon cancer histology images. IEEE transactions on medical imaging 35, 1196-1206 (2016)

10. Basha, S.S., Ghosh, S., Babu, K.K., Dubey, S.R., Pulabaigari, V., Mukherjee, S.: Rccnet: An efficient convolutional neural network for histological routine colon cancer nuclei classification. In: 2018 15th International Conference on Control, Automation, Robotics and Vision (ICARCV), pp. 1222-1227. IEEE, (Year)

11. Gamper, J., Alemi Koohbanani, N., Benet, K., Khuram, A., Rajpoot, N.: Pannuke: an open pan-cancer histology dataset for nuclei instance segmentation and classification. In: European Congress on Digital Pathology, pp. 11-19. Springer, (Year)

12. Hou, L., Agarwal, A., Samaras, D., Kurc, T.M., Gupta, R.R., Saltz, J.H.: Robust histopathology image analysis: To label or to synthesize? In: Proceedings of the IEEE/CVF Conference on Computer Vision and Pattern Recognition, pp. 8533-8542. (Year)

13. Gong, X., Chen, S., Zhang, B., Doermann, D.: Style consistent image generation for nuclei instance segmentation. In: Proceedings of the IEEE/CVF Winter Conference on Applications of Computer Vision, pp. 3994-4003. (Year)

14. Zhang, H., Cisse, M., Dauphin, Y.N., Lopez-Paz, D.: mixup: Beyond empirical risk minimization. arXiv preprint arXiv:1710.09412 (2017)

15. DeVries, T., Taylor, G.W.: Improved regularization of convolutional neural networks with cutout. arXiv preprint arXiv:1708.04552 (2017)

16. Yun, S., Han, D., Oh, S.J., Chun, S., Choe, J., Yoo, Y.: Cutmix: Regularization strategy to train strong classifiers with localizable features. In: Proceedings of the IEEE/CVF international conference on computer vision, pp. 6023-6032. (Year)

17. Telea, A.: An image inpainting technique based on the fast marching method. Journal of graphics tools 9, 23-34 (2004)

18. Doan, T.N.N., Song, B., Le Vuong, T.T., Kim, K., Kwak, J.T.: SONNET: A self-guided ordinal regression neural network for segmentation and classification of nuclei in large-scale multi-tissue histology images. IEEE Journal of Biomedical and Health Informatics (2022)

19. Tan, M., Le, Q.: Efficientnet: Rethinking model scaling for convolutional neural networks. In: International conference on machine learning, pp. 6105-6114. PMLR, (Year)